\documentclass[aps,pra,floatfix,showpacs,notitlepage,twocolumn]{revtex4-1}

\usepackage{verbatim}
\usepackage{dcolumn}
\usepackage{mathrsfs}
\usepackage{amsmath}
\usepackage{amssymb}
\usepackage{bm}
\usepackage{graphicx}
\usepackage{subfigure}

\begin{document}


\title{Rank Restriction for the Variational Calculation of Two-electron Reduced Density Matrices of Many-electron Atoms and Molecules}

\author{Kasra Naftchi-Ardebili}
\author{Nathania W. Hau}
\author{David A. Mazziotti}

\email{damazz@uchicago.edu}

\affiliation{Department of Chemistry and The James Franck Institute, The University of Chicago, Chicago, IL 60637}%

\date{Submitted September 13, 2011; Published Phys. Rev. A {\bf 84}, 052506 (2011)}



\begin{abstract}

Variational minimization of the ground-state energy as a function of
the two-electron reduced density matrix (2-RDM), constrained by
necessary $N$-representability conditions, provides a
polynomial-scaling approach to studying strongly correlated
molecules without computing the many-electron wavefunction.  Here we
introduce a new route to enhancing necessary conditions for
$N$-representability through rank restriction of the 2-RDM.  Rather
than adding computationally more expensive $N$-representability
conditions, we directly enhance the accuracy of two-particle
(2-positivity) conditions through rank restriction, which removes
degrees of freedom in the 2-RDM that are not sufficiently
constrained. We select the rank of the particle-hole 2-RDM by
deriving the ranks associated with model wavefunctions including
both mean-field and antisymmetrized geminal power (AGP) wave
functions. Because the 2-positivity conditions are exact for quantum
systems with AGP ground states, the rank of the particle-hole 2-RDM
from the AGP ansatz provides a minimum for its value in variational
2-RDM calculations of general quantum systems.  To implement the
rank-restricted conditions, we extend a first-order algorithm for
large-scale semidefinite programming.  The rank-restricted
conditions significantly improve the accuracy of the energies; for
example, the percentages of correlation energies recovered for HF,
CO, and N$_{2}$ improve from 115.2\%, 121.7\%, and 121.5\% without
rank restriction to 97.8\%, 101.1\%, and 100.0\% with rank
restriction.  Similar results are found at both equilibrium and
non-equilibrium geometries.  While more accurate, the
rank-restricted $N$-representability conditions are less expensive
computationally than the full-rank conditions.

\end{abstract}

\maketitle

\section{Introduction}

Significantly more information is encoded within the wavefunction
than is necessary for the calculation of energies and properties of
many-electron quantum systems. In 1955 Mayer proposed in {\em
Physical Review} calculating the ground-state energy variationally
as a functional of the two-electron reduced density matrix (2-RDM)
rather than the $N$-electron wavefunction~\cite{RDM,CY00,M55}.
Unlike the wavefunction the 2-RDM scales polynomially with the
number $N$ of electrons in the molecule.  In further work, however,
it became apparent that the 2-RDM must be constrained by non-trivial
conditions to ensure that it is representable by an $N$-electron
density matrix (or wavefunction), and the search for these
conditions became known as the {\em N-representability
problem}~\cite{CY00,C63,GP64,E78,H79,EJ00a,ME01,NNE01,M02,ZBF04,M04c,
M04b,CSL06,M06b,M06c,M07b,NBF08,BPZ07,GM08a,VVV09,GM10b,SI10}.  For
nearly 50 years the direct calculation of the 2-RDM without the
wavefunction was stymied by the need for better $N$-representability
conditions and better optimization methods.

The variational computation of an $N$-particle system's ground-state
energy as a functional of the 2-RDM has recently been realized
through advances in ({\em i}) developing $N$-representability
conditions~\cite{E78,EJ00a,ME01,M02,ZBF04,VVV09} and ({\em ii})
designing optimization algorithms~\cite{ZBF04,M04c,M04b,CSL06,E07,
M07e,FNY07,M11}. A systematic hierarchy of $N$-representability
conditions has been developed in the form of {\em $p$-positivity
conditions}~\cite{ME01,M02,M06c,GP64}, which constrain $p+1$
distinct metric matrices of the fermionic $p$-RDM to be positive
semidefinite (A matrix is {\em positive semidefinite} if and only if
its eigenvalues are nonnegative). The $p+1$ metric matrices
represent the probability distributions of $p-q$ particles and $q$
holes with $q$ ranging from $0$ to $p$ where a hole is the absence
of a particle~\cite{ME01}. The $p$-positivity conditions ensure that
each of these probability distributions is nonnegative. These
conditions, even for small $p$, are capable of capturing both
moderate and strong electron correlation; for example, the
2-positivity conditions are necessary and sufficient for computing
the ground-state energies of pairing Hamiltonians~\cite{M02}, often
employed in modeling long-range order and superconductivity.

Minimizing the ground-state energy as a 2-RDM functional constrained
by these conditions produces a special type of optimization known as
{\em semidefinite programming} (SDP)~\cite{E79,VB96,HSV00,EJ00b,
BuMo03-1,M04c,M04b,CSL06,E07,M07e,FNY07,M11}.  Importantly, because
SDP problems are solvable in polynomial time, the variational 2-RDM
method provides a {\em polynomial-time relaxation} of the
exponentially scaling many-electron problem that is suitable for
describing strong electron correlation.  The 2-RDM methodology has
been applied in quantum chemistry and condensed-matter physics to
studying many-electron molecules and their reactions~\cite{GM08a,
GM09b,GM10b,PGG11}, quantum phase transitions~\cite{GM06b,SM09},
quantum dots~\cite{RM08b}, molecular clusters~\cite{KM07a,KM09} and
spin systems like the Hubbard~\cite{HM06a} and Ising~\cite{SGM10}
models.   While new wavefunction methods for strong correlation are
being actively developed~\cite{KS03,
MA03,LMH09,MAG11,GS09a,THS10,DMRG,PLH09}, traditional wavefunction
methods are limited to linear combinations of approximately $10^{9}$
molecular configurations.

In this article we present a new approach to improving the accuracy
of energies from the 2-RDM method through rank restriction of the
$N$-representability conditions.  The accuracy of the 2-RDM
calculations with 2-positivity can be dramatically enhanced through
the addition of 3-positivity conditions~\cite{ZBF04,M05,GM07a,M06c},
but these conditions significantly increase the total computational
cost of the calculations.  Rather than turning to 3-positivity, we
propose to improve 2-positivity more directly without an increase in
its computational cost.  The central idea is that {\em approximate
$N$-representability conditions like the 2-positivity constraints
should be combined with less flexibility in the 2-RDM than more
stringent $N$-representability conditions like 3-positivity.}

One systematic approach to controlling the flexibility of the 2-RDM
is to restrict its rank or the rank of one of its metric matrices.
We can motivate the selection of the rank by examining the ranks
associated with model wavefunctions including mean-field~\cite{J07}
and antisymmetrized geminal power (AGP)~\cite{CY00,C65,L80,C97,R00,
GFT00,GFT01,CAS04,BMD06,SJH11} wavefunctions. Importantly, because
the 2-positivity conditions yield the exact ground-state energy of
any AGP Hamiltonian---that is, a Hamiltonian whose ground state is
described by an AGP wavefunction~\cite{CY00,M02}, the AGP 2-RDM
provides a lower bound on the optimal rank of the particle-hole form
of the 2-RDM. The resulting rank-restricted $N$-representability
conditions yield significantly improved ground-state energies at a
slightly lower computational cost than unrestricted 2-positivity
conditions.

After theoretical results are presented in section~\ref{sec:the},
illustrative applications are made in section~\ref{sec:app} to
computing ground-state energies for a set of molecules in several
basis sets as well as bond stretching of hydrogen fluoride and
diatomic nitrogen. Section~\ref{sec:con} provides a brief discussion
and concluding remarks.

\section{Theory}

\label{sec:the}

The energy is expressed as a functional of the 2-RDM in
sections~\ref{sec:en}, and the $N$-representability constraints,
known as 2-positivity conditions~\cite{GP64,ME01}, are reviewed in
sections~\ref{sec:pos}.  In section~\ref{sec:wf} we derive the
maximum rank of the particle-hole ${}^{2} G$ matrix for two model
wavefunctions, the Hartree-Fock wavefunction and the AGP
wavefunction. For the AGP wavefunction the maximum rank for each
block of the spin-adapted particle-hole $^{2} G$ matrix is also
derived. Finally, in section~\ref{sec:sdp} we extend a large-scale
algorithm for SDP~\cite{M04c,M07e} to support rank restriction.

\subsection{Energy functional}

\label{sec:en}

Because electrons are indistinguishable with pairwise interactions,
the energy of any $N$-electron quantum system can be expressed as a
{\em linear} functional of the two-electron reduced Hamiltonian
matrix $^{2} K$ and the two-electron reduced density matrix
(2-RDM)~\cite{RDM,CY00,M55}
\begin{eqnarray}
\label{eq:E} E & = & \sum_{p,q,s,t}{{}^{2} K^{p,q}_{s,t}
\hspace{1mm} {}^{2}
D^{p,q}_{s,t}} \\
E & = & {\rm Tr}({}^{2} K \hspace{1mm} {}^{2} D) ,
\end{eqnarray}
where the indices denote spin orbitals $\phi_{p}$ in a finite
one-electron basis set $\{ \phi_{p} \}$.  The elements of the
two-electron reduced Hamiltonian matrix are
\begin{equation}
\label{eq:K2} {}^{2} K^{p,q}_{s,t} = \frac{1}{N-1}{}^{1} K^{p}_{s}
\delta^{q}_{t} + {}^{2} V^{p,q}_{s,t} ,
\end{equation}
where matrices ${}^{1} K$ and ${}^{2} V$ contain the one- and
two-electron integrals respectively, and the elements of the 2-RDM
are
\begin{equation}
{}^{2} D^{p,q}_{s,t} = \langle \Psi | {\hat a}^{\dagger}_{p} {\hat
a}^{\dagger}_{q} {\hat a}_{t} {\hat a}_{s} | \Psi \rangle
\end{equation}
where ${\hat a}^{\dagger}_{p}$ $({\hat a}_{p})$ is a creation
(annihilation) operator in second quantization that creates
(annihilates) an electron in spin orbital $\phi_{p}$ and $\Psi$
represents the $N$-electron wavefunction.

\subsection{$N$-representability conditions}

\label{sec:pos}

Because not every two-electron density matrix is representable by an
$N$-electron density matrix, the 2-RDM must be constrained by
$N$-representability conditions~\cite{CY00,C63,GP64,E78,EJ00a,
ME01,NNE01,M02,ZBF04,M04c,M04b,CSL06,M06b,M06c,M07b,NBF08,BPZ07,
GM08a,VVV09,GM10b,SI10}. A systematic hierarchy of constraints is
furnished by the $p$-positivity conditions~\cite{ME01,M02,M06c,
GP64}. The {\em 1-positivity conditions}, constraining the
one-particle ${}^{1} D$ and the one-hole ${}^{1} Q$ RDMs to be
positive semidefinite, corresponds to restricting the eigenvalues of
the 1-RDM $n_{p}$, known as natural occupation numbers, to lie
between zero and one $n_{p} \in [0,1]$, which enforces the {\em
Pauli principle}. Coleman~\cite{C63,CY00} first proved that these
relatively simple conditions plus the usual trace, Hermiticity, and
antisymmetry constraints in the definition of a density matrix are
not only necessary but also {\em sufficient} for the 1-RDM to
represent an $N$-electron density matrix.

The {\em 2-positivity conditions}~\cite{ME01,GP64}, providing
necessary constraints on the 2-RDM, constrain the following three
metric matrices to be positive semidefinite:
\begin{eqnarray}
{}^{2} D & \succeq & 0  \label{eq:D} \\
{}^{2} Q & \succeq & 0  \label{eq:Q} \\
{}^{2} G & \succeq & 0, \label{eq:G}
\end{eqnarray}
where the metric matrices ${}^{2} D$, ${}^{2} Q$, and ${}^{2} G$
correspond to the probability distributions for two particles, two
holes, and one particle and one hole.  In second quantization the
elements of these matrices are expressible as
\begin{equation}
\label{eq:X} {}^{2} X^{p,q}_{s,t} = \langle \Psi | ^{X} {\hat
C}_{p,q} \hspace{1mm} ^{X} {\hat C}_{s,t}^{\dagger} | \Psi \rangle,
\end{equation}
where
\begin{eqnarray}
^{D} {\hat C}_{p,q} & = & {\hat a}_{p}^{\dagger} {\hat a}_{q}^{\dagger} \\
^{Q} {\hat C}_{p,q} & = & {\hat a}_{p} {\hat a}_{q} \\
^{G} {\hat C}_{p,q} & = & {\hat a}_{p}^{\dagger} {\hat a}_{q} .
\label{eq:CG}
\end{eqnarray}
All three metric matrices contain equivalent information in the
sense that rearranging the creation and annihilation operators
produces linear mappings between the elements of the three
matrices~\cite{RDM,CY00,M02}; particularly, the 2-hole RDM ${}^{2}
Q$ and the particle-hole RDM ${}^{2} G$ can be written in terms of
the 2-particle RDM ${}^{2} D$ as follows
\begin{equation}
\label{eq:Q2} {}^{2} Q^{p,q}_{s,t} = 2 \hspace{1mm} {}^{2}
I^{p,q}_{s,t} - 4 \hspace{1mm} {}^{1} D^{p}_{s} \wedge {}^{1}
I^{q}_{t} + {}^{2} D^{p,q}_{s,t}
\end{equation}
and
\begin{equation}
\label{eq:G2} {}^{2} G^{p,q}_{s,t} = {}^{1} I^{q}_{t} \hspace{1mm}
{}^{1} D^{p}_{s} - {}^{2} D^{p,t}_{s,q} ,
\end{equation}
where ${}^{1} I$ and ${}^{2} I$ are the one- and two-particle
identity matrices and $\wedge$ denotes the Grassmann wedge
product~\cite{M98a,M98b}.  While all three matrices are
interconvertible, the nonnegativity of the eigenvalues of one matrix
does not imply the nonnegativity of the eigenvalues of the other
matrices, and hence, each semidefinite constraint in
Eqs.~(\ref{eq:D}), (\ref{eq:Q}), and (\ref{eq:G}) provides an
important $N$-representability condition.

\subsection{Rank restriction}

\subsubsection{Model wavefunctions}

\label{sec:wf}

The best known model wavefunction is the mean-field wavefunction
introduced by Hartree, Fock, and Slater~\cite{J07}.  In first
quantization the $N$-electron {\em Hartree-Fock} wavefunction can be
expressed as
\begin{equation}
\Psi_{\rm HF} = \phi_{1}(1) \wedge \phi_{2}(2) \wedge ... \wedge
\phi_{N}(N),
\end{equation}
while in second quantization it can be written as
\begin{equation}
| \Psi_{\rm HF} \rangle = \left ( \prod_{i=1}^{N}{ {\hat
a}^{\dagger}_{i} } \right ) | 0 \rangle,
\end{equation}
where $| 0 \rangle$ denotes the vacuum state, the state without any
electrons.  The rank of the particle-hole 2-RDM (or ${}^{2} G$),
whose elements are given in Eq.~(\ref{eq:X}) equals the number of
linearly independent $N$-electron functions $ | f_{i,j} \rangle$
having the form
\begin{equation}
\label{eq:f} | f_{i,j} \rangle = {\hat a}^{\dagger}_{j} {\hat a}_{i}
| \Psi \rangle .
\end{equation}
For the Hartree-Fock wavefunction the set $\{ | f_{i,j} \rangle \}$
contains the wavefunction $ | \Psi_{\rm HF} \rangle$ itself as well
as $(r-N)N$ functions from all single excitations of $ | \Psi_{\rm
HF} \rangle$. Hence, the rank of ${}^{2} G$ from a Hartree-Fock
wavefunction is $(r-N)N+1$. For a wavefunction to describe a
correlated $N$-electron system in $r$ spin orbitals its
particle-hole 2-RDM ${}^{2} G$ must have a rank strictly larger than
$(r-N)N+1$.

A flexible model wavefunction with electron correlation is the {\em
antisymmetrized geminal power} (AGP) wavefunction~\cite{CY00,C65,
L80,C97,R00,GFT00,GFT01,CAS04,BMD06,SJH11}, also known as the
projected Bardeen, Cooper, and Schrieffer (BCS) wavefunction, which
can be employed to model Cooper pairing in superconductivity. The
$N$-electron AGP wavefunction in first quantization can be written
as
\begin{equation}
\Psi_{\rm AGP} = g(1,2) \wedge g(3,4) \wedge ... \wedge g(N-1,N),
\end{equation}
where $g(1,2)$ is a {\em two-electron} function (or geminal) in
contrast to the set of one-electron orbitals $\{ \phi_{i} \}$.  In
second quantization we can define the AGP wavefunction as a
projection of the BCS wavefunction onto the $N$-electron space
\begin{equation}
\label{eq:bcs} | \Psi_{\rm AGP} \rangle = {\hat P}_{N} \left [
\prod_{i=1}^{r/2}{ ( 1 + \gamma_{i} {\hat a}^{\dagger}_{+i} {\hat
a}^{\dagger}_{-i} ) } \right ] | 0 \rangle,
\end{equation}
where $| 0 \rangle$ is the vacuum state and ${\hat P}_{N}$ is the
projection operator that projects the BCS wavefunction onto the
Hilbert space of $N$-electron wavefunctions.  A key feature of the
AGP wavefunction is the special pairing of orbitals~\cite{CY00},
which we denote by $+i$ and $-i$ for $i \in [1,r/2]$.  In the study
of superconductivity this pairing is employed to model the observed
Cooper pairing of the momenta of electrons.

The rank of ${}^{2} G$, again equaling the number of linearly
independent $N$-electron functions $| f_{i,j} \rangle$ in
Eq.~(\ref{eq:f}), can be determined for AGP from the pairing of
orbitals.  For AGP the functions $| f_{i,j} \rangle$ can be divided
into two classes~\cite{R00}:
\begin{eqnarray}
| f_{i,i}^{P} \rangle & = & {\hat P}_{i,i} | \Psi_{\rm AGP} \rangle \\
| f_{i,j}^{Q} \rangle & = & {\hat Q}_{i,j} | \Psi_{\rm AGP} \rangle,
\end{eqnarray}
where the ${\hat P}_{i,i}$ are projection operators and ${\hat
Q}_{i,j}$ are operators whose adjoint operators annihilate the AGP
wavefunction, that is
\begin{equation}
{\hat Q}_{i,j}^{\dagger} | \Psi_{\rm AGP} \rangle = 0.
\end{equation}
Specifically, when the $\gamma_{i}$ are not more than doubly
degenerate, $r/2$ linearly independent functions $| f_{i,i}^{P}
\rangle$ arise from the projectors
\begin{equation}
\label{eq:P} {\hat P}_{i,i} = {\hat a}^{\dagger}_{i} {\hat a}_{i} ,
\end{equation}
and $r(r-2)/2$ linearly independent functions $| f_{i,j}^{Q}
\rangle$ arise from the ${\hat Q}_{i,j}$ operators whose adjoints
are
\begin{equation}
\label{eq:Qd} {\hat Q}_{i,j}^{\dagger} = \gamma_{i} {\hat
a}^{\dagger}_{i} {\hat a}_{j} - {\rm sign}(ij) \gamma_{j} {\hat
a}^{\dagger}_{-j} {\hat a}_{-i},
\end{equation}
where $i,j \in [-r/2,r/2]\backslash[0]$ with $i \neq j$ and $i \neq
-j$ and ${\rm sign}(ij)$ returns the sign of the product of $i$ and
$j$.

The fact that each of the $r(r-2)/2$ operators ${\hat
Q}_{i,j}^{\dagger}$ annihilates the AGP wavefunction follows from
the pairing property of the orbitals~\cite{CY00,R00}.  From the
definition of the AGP wavefunction in Eq.~(\ref{eq:bcs}), it can be
seen that in each Slater determinant contributing to the AGP
wavefunction both orbitals in a pair, i.e. $\phi_{+i}$ and
$\phi_{-i}$, are either occupied or unoccupied.  Furthermore, each
pair in the wavefunction is weighted by a corresponding element of
the vector $\gamma$. Hence, the actions of the operators $\gamma_{i}
{\hat a}^{\dagger}_{i} {\hat a}_{j}$ and $\gamma_{j} {\hat
a}^{\dagger}_{-j} {\hat a}_{-i}$ on the AGP wavefunction are always
equal or opposite in sign depending on whether the function ${\rm
sign}(i,j)$ is equal to +1 or -1, which proves the result. The
numbers of linearly independent $| f_{i,i}^{P} \rangle$ and $|
f_{i,j}^{Q} \rangle$ will be less than their maximum values of $r/2$
and $r(r-2)/2$ if the $\gamma_{i}$ where $\gamma_{i}=\gamma_{-i}$
are more than doubly degenerate. Such a case occurs when the AGP
wavefunction reduces to the Hartree-Fock wavefunction and the
numbers of linearly independent $| f_{i,i}^{P} \rangle$ and $|
f_{i,j}^{Q} \rangle$ become 1 and $N(r-N)$, respectively.
Consequently, for an AGP wavefunction the maximum rank of $^{2} G$
is $r(r-2)/2+r/2$ or $r(r-1)/2$.

In electronic calculations, when the expectation value of the
$z$-component of the spin operator $\langle {\hat S}_{z} \rangle$
vanishes, the basis functions of the ${}^{2} G$ metric matrix can be
{\em spin adapted} to produce a block diagonal ${}^{2} G$ matrix
with four blocks~\cite{GM05c}.  The four blocks correspond to the
following four $^{G} {\hat C}_{{\bar p},{\bar q}}$ operators:
\begin{eqnarray}
^{G} {\hat C}_{{\bar i},{\bar j}}^{(0,0)} & = &
\frac{1}{\sqrt{2}}(\hat{a}_{{\bar i}\alpha}^{\dagger}\hat{a}_{{\bar
j}\alpha} +\hat{a}_{{\bar i}\beta}^{\dagger}\hat{a}_{{\bar j}\beta}) \\
^{G} {\hat C}_{{\bar i},{\bar j}}^{(1,-1)} & = &
\hat{a}^{\dagger}_{{\bar i}\beta} \hat{a}_{{\bar j}\alpha} \\
^{G} {\hat C}_{{\bar i},{\bar j}}^{(1,0)} & = &
\frac{1}{\sqrt{2}}(\hat{a}_{{\bar i}\alpha}^{\dagger} \hat{a}_{{\bar
j}\alpha} - \hat{a}_{{\bar i}\beta}^{\dagger} \hat{a}_{{\bar j}\beta}) \\
^{G} {\hat C}_{{\bar i},{\bar j}}^{(1,+1)} & = &
\hat{a}^{\dagger}_{{\bar i}\alpha} \hat{a}_{{\bar j}\beta},
\end{eqnarray}
where the bar above the index refers to the {\em spatial} part of
the orbital, the spin part of each orbital is denoted as either
$\alpha$ ($+1/2$) or $\beta$ ($-1/2$), and the upper right indices
of ${^{G} {\hat C}}_{{\bar i},{\bar j}}^{s,m}$ denote the square of
the total spin and the $z$-component of the total spin for the
two-electron operators.  If the pairing within the AGP ansatz is
taken to be between spin orbitals sharing the same spatial
component, the AGP wavefunction in Eq.~(\ref{eq:bcs}) can be
re-written with ${\bar i}\alpha$ and ${\bar i}\beta$ replacing $+i$
and $-i$.

To determine the rank of the ${}^{2} G$ spin blocks, we can spin
adapt the projection operators in Eq.~(\ref{eq:P}) and the adjoint
of the annihilation operators in Eq.~(\ref{eq:Qd}), respectively, to
obtain
\begin{equation}
{\hat P}_{{\bar i},{\bar i}}^{(0,0)} = \frac{1}{\sqrt{2}} ( {\hat
P}_{{\bar i}\alpha,{\bar i}\alpha} + {\hat P}_{{\bar i}\beta,{\bar
i}\beta} )
\end{equation}
and
\begin{eqnarray}
{\hat Q}_{{\bar i},{\bar j}}^{(0,0)}  & = & \frac{1}{\sqrt{2}} (
{\hat Q}_{{\bar i}\alpha,{\bar j}\alpha} + {\hat Q}_{{\bar i}\beta,{\bar j}\beta} ), \\
{\hat Q}_{{\bar i},{\bar j}}^{(1,-1)} & = & {\hat Q}_{{\bar i}\alpha,{\bar j}\beta}, \\
{\hat Q}_{{\bar i},{\bar j}}^{(1,0)}  & = & \frac{1}{\sqrt{2}} (
{\hat Q}_{{\bar i}\alpha,{\bar j}\alpha} - {\hat Q}_{{\bar i}\beta,{\bar j}\beta} ), \\
{\hat Q}_{{\bar i},{\bar j}}^{(1,+1)} & = & {\hat Q}_{{\bar
i}\beta,{\bar j}\alpha}.
\end{eqnarray}
All $r_{s}$ spin-adapted projection operators contribute to the
$(0,0)$ spin block of ${}^{2} G$, and $r_{s} ( r_{s}-1 )/2$ ${\hat
Q}$-type operators contribute to each of the $(0,0)$, $(1,-1)$,
$(1,0)$, and $(1,+1)$ spin blocks where the number $r_{s}$ of
spatial orbitals equals one-half the number $r$ of spin orbitals.
Hence, the rank of the singlet spin block $(0,0)$ of ${}^{2} G$ is
$r_{s} ( r_{s}+1 )/2$, and the ranks of the three triplet spin
blocks of ${}^{2} G$ are $r_{s} ( r_{s}-1 )/2$.  When $\langle {\hat
S}_{z} \rangle = 0$, all three triplet blocks are
identical~\cite{GM05c}.

\subsubsection{Semidefinite programming}

\label{sec:sdp}

The variational 2-RDM method with 2-positivity conditions minimizes
the ground-state energy as a 2-RDM functional
\begin{equation}
\label{eq:Ex} {\rm minimize~~}  E(x) = c^{T} x
\end{equation}
where the vector $c$ contains information about the quantum system
in the form of the two-electron reduced Hamiltonian in
Eq.~(\ref{eq:K2})~\cite{ME01} and the vector $x$ contains the three
different metric-matrix forms of the 2-RDM whose elements are given
in Eq.~(\ref{eq:X}).  Because the three metric matrices in $x$ are
{\em interrelated} by linear mappings
\begin{equation}
\label{eq:Ax} A x = b
\end{equation}
and {\em constrained} to be positive semidefinite
\begin{equation}
\label{eq:Mx} M(x) = \left (
\begin{array}{ccc}
       {}^{2} D & 0 & 0 \\
       0 & {}^{2} Q & 0 \\
       0 & 0 & {}^{2} G
\end{array} \right ) \succeq 0 ,
\end{equation}
where the operator $M$ maps the vector $x$ to a matrix, the energy
minimization constitutes a special type of constrained optimization
known as semidefinite programming (SDP)~\cite{E79,VB96,HSV00,
EJ00b,BuMo03-1,M04c,M04b,CSL06,E07,M07e,FNY07,M11}.  SDP is a
generalization of linear programming from linear scalar inequalities
to linear matrix inequalities.

Second-order algorithms for SDP, developed in the 1990s~\cite{VB96,
HSV00}, have an expensive $r^{16}$ scaling~\cite{NNE01,M02} in
floating-point operations when applied to variational 2-RDM
calculations with 2-positivity constraints. Zhao {\em et
al.}~\cite{ZBF04} introduced a dual formulation of the 2-RDM
optimization that decreased the computational scaling to $r^{12}$,
and one of the authors developed two first-order algorithms, a
matrix-factorization method~\cite{M04b,M04c,M07e} and a
boundary-point method~\cite{M11}, that reduce the floating-point
operations to $r^{6}$ and the memory requirements from $r^{8}$ to
$r^{4}$.  Canc{\'e}s, Stoltz, and Lewin~\cite{CSL06}, who studied a
dual formulation of the SDP problem, confirmed the efficiency of the
matrix factorization method, and Verstichel {\em et
al.}~\cite{VVV09} introduced a first-order algorithm, based on
interior-point methods.

For the rank-restricted $N$-representability conditions the SDP
optimization must be modified to include rank restriction of the
particle-hole ${}^{2} G$ metric matrix within $M$.  In the
matrix-factorization method the solution matrix $M$ is explicitly
constrained to be positive semidefinite by a matrix
factorization~\cite{M04b,M04c,M07e}:
\begin{equation}
\label{eq:RR} M = R R^{*} .
\end{equation}
Importantly, the rank of $M$ or any of its subblocks can be readily
constrained to an integer $q$ by restricting the number of columns
of $R$ to $q$ where $q$ is less than the dimension of the square
matrix $M$.  With this flexibility we can solve SDP problems both
with and without rank restriction.  If the rank of a block in $R$ is
restricted to an unphysical value such as an integer less than the
rank correspond to a Hartree-Fock model wavefunction, the algorithm
generally will not converge.  Otherwise, convergence of the
rank-restricted SDP is similar to that of the unrestricted SDP.

\section{Applications}

\label{sec:app}

After an overview of computational details and a summary of
$N$-representability conditions, we present results of the
rank-restricted variational 2-RDM method for molecules at both
equilibrium and non-equilibrium geometries.

\subsection{Computational details}

The variational 2-RDM method with 2-positivity and rank-restricted
2-positivity conditions is illustrated with calculations on several
molecules at equilibrium and non-equilibrium geometries in minimal
Slater-type orbital (STO-6G)~\cite{sto}, double-zeta (DZ)~\cite{dz},
and correlation-consistent polarized double-zeta (cc-pVDZ)~\cite{pz}
basis sets. Non-equilibrium geometries are obtained from the {\it
Handbook of Chemistry and Physics}~\cite{crc}, all core orbitals are
double occupied (frozen), and the molecules are in singlet states.
The calculation of one- and two-electron integrals and full
configuration interaction (FCI) is implemented in the quantum
chemistry package GAMESS (USA)~\cite{gamess}.

\subsection{Summary of \textit{N}-representability conditions}

Variational RDM ground-state energies are computed with the
first-order nonlinear SDP algorithm developed by
Mazziotti~\cite{M04b,M04c,M07e}. The following
\textit{N}-representability conditions are enforced:

\textbf{(1)} Hermiticity of the 2-RDM:
\begin{equation}
\label{eqn:hermiticity}
{}^{2}D^{i,j}_{k,l}={}^{2}D^{k,l}_{i,j}\hspace{1mm}.
\end{equation}

\textbf{(2)} Antisymmetry of upper and lower indices
\begin{equation}
\label{eqn:antisymmetry}
{}^{2}D^{i,j}_{k,l}=-\hspace{1mm}^{2}D^{j,i}_{k,l}=-\hspace{1mm}^{2}D^{i,j}_{l,k}=\hspace{1mm}^{2}D^{j,i}_{l,k}\hspace{1mm},
\end{equation}
is enforced by antisymmetrized basis functions
$\tilde{\phi}_{i,j}=1/\sqrt{2}(\phi_{i,j}-\phi_{j,i})$.

\textbf{(3)} Trace conditions on the spin-adapted blocks of the
2-RDM~\cite{GM05c}:
\begin{eqnarray}
\label{eqn:spinadapttrace}
\begin{array}{ll}
\text{Tr}\left(\hspace{1mm}^{2}D^{(1,0)}\right) & = N_{s}(N_{s}-1) \\
\text{Tr}\left(\hspace{1mm}^{2}D^{(0,0)}\right) & = N_{s}(N_{s}+1)
\hspace{1mm} ,
\end{array}
\end{eqnarray}
where $N_{s}=N/2$.

\textbf{(4)} Contraction of the spin-adapted 2-RDM~\cite{GM05c} onto
the 1-RDM:
\begin{eqnarray}
\label{eqn:spinadaptcontr}
\begin{array}{ll}
(N_{s}-1)\hspace{1mm}^{1} D^{i\alpha}_{j\alpha} & =
\sum_{k}\hspace{1mm}^{2} D^{(1,1)}_{i,k;j,k} \\
(N_{s}+1)\hspace{1mm}^{1} D^{i\alpha}_{j\alpha} & =
\sum_{k}\hspace{1mm}^{2} D^{(0,0)}_{i,k;j,k}
\end{array}
\end{eqnarray}

\textbf{(5)} The 2-positivity conditions [Eqs.~(\ref{eq:D}-
\ref{eq:G})], on three different representations of the 2-RDM whose
elements, given in Eq.~(\ref{eq:X}), are related by the linear
mappings in Eqs.~(\ref{eq:Q2}) and~(\ref{eq:G2}).

\textbf{(6)} In the case of rank restriction, the rank of the
particle-hole matrix ${}^{2} G^{(0,0)}$ is restricted.

\subsection{Results}

\begin{figure}[t!]

\includegraphics[scale=0.4]{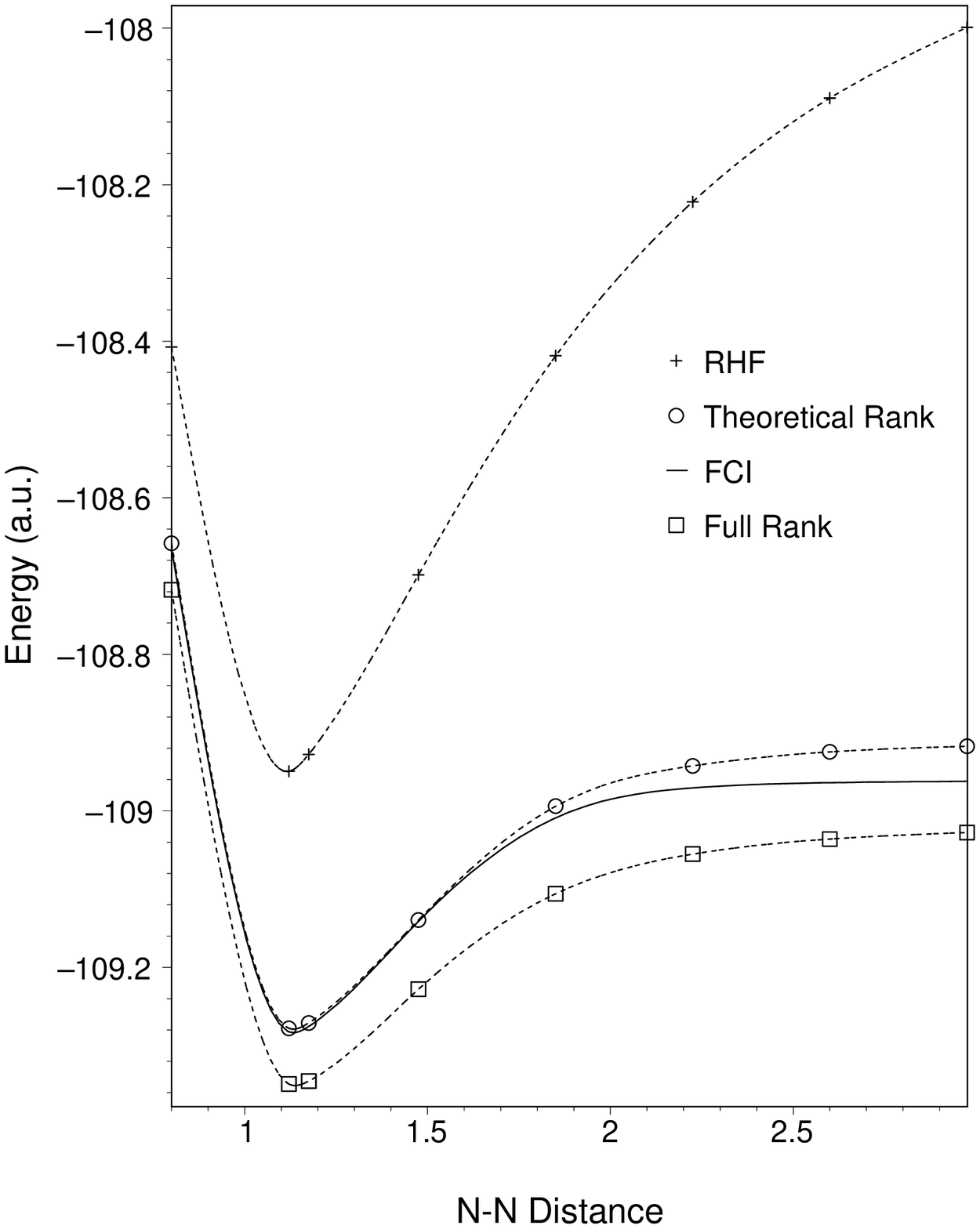}

\caption{The potential energy curve of the nitrogen molecule N$_{2}$
in the cc-pVDZ basis set from the variational 2-RDM method with and
without rank restriction.  Results are compare with those from
Hartree-Fock and FCI.}

\label{f:n2}

\end{figure}

\begin{figure}[htp!]

\includegraphics[scale=0.4]{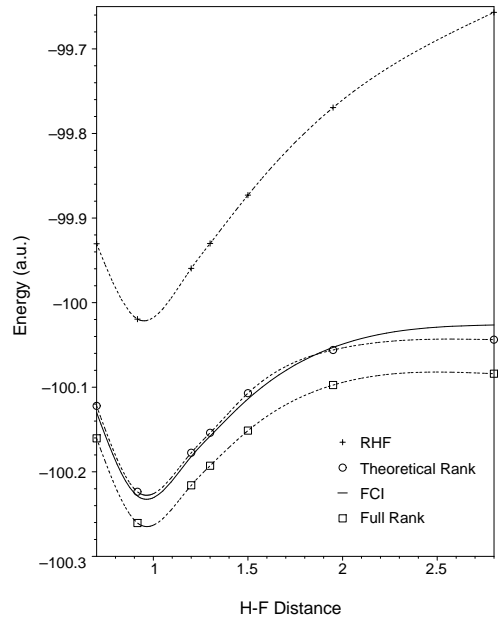}

\caption{The potential energy curve of the hydrogen fluoride
molecule in the cc-pVDZ basis set from the variational 2-RDM method
with and without rank restriction. Results are compare with those
from Hartree-Fock and FCI.}

\label{f:hf}

\end{figure}

Two sets of \textit{N}-representability constraints are imposed in
the calculations shown in Tables~\ref{t:mol}-\ref{t:hf} and
Figs.~\ref{f:n2} and~\ref{f:hf}: ({\em i}) 2-positivity conditions
without rank restriction, labeled {\em full rank}, ({\em ii})
2-positivity conditions plus rank restriction, labeled {\em
theoretical rank}, in which the rank of the ${}^{2} G^{(0,0)}$ block
of the particle-hole matrix is restricted to
$r_{s}(r_{s}+1)/2$---its maximum value from a model AGP
wavefunction.

\begin{table*}[htp!]

\caption{The percentage of the correlation energy recovered by the
variational 2-RDM method with full-rank and rank-restricted
$N$-representability conditions is shown for the molecules CO,
N$_{2}$, H$_{2}$O, HF, and NO$^+$ in a variety of basis sets.}

\label{t:mol}

\begin{ruledtabular}
\begin{tabular}{cccccccc}
\multicolumn{2}{c}{} & \multicolumn{1}{c}{Full} &
\multicolumn{1}{c}{Theoretical} & \multicolumn{1}{c}{Full CI} &
\multicolumn{1}{c}{Correlation} &
\multicolumn{2}{c}{\% Correlation Energy} \\
\cline{7-8} Molecule & Basis Set & \multicolumn{1}{c}{Rank} &
\multicolumn{1}{c}{Rank} & \multicolumn{1}{c}{Energy (a.u.)} &
\multicolumn{1}{c}{Energy (a.u.)} & \multicolumn{1}{c}{Full Rank} &
\multicolumn{1}{c}{Theoretical Rank} \\
\hline
CO &  STO-6G & 64 & 36 & -112.443174 & -0.139676 & 108.5 & 101.2 \\
 &  DZ & 256 & 136 & -112.893590 & -0.208672 & 119.3 & 102.5 \\
 & cc-pVDZ & 676 & 351 & -113.054884 & -0.305767 & 121.7 & 101.1 \\

N$_{2}$ & STO-6G & 64 & 36 & -108.699813 & -0.158189 & 107.6 & 97.7 \\
 & DZ & 256 & 136 & -109.104172 & -0.226029 & 119.1 & 102.8 \\
 & cc-pVDZ & 676 & 351 & -109.278339 & -0.329000 & 121.5 & 100.0 \\

H$_{2}$O & STO-6G & 36 & 21 & -75.728838 & -0.050041 & 104.0 & 101.7 \\
 & DZ & 144 & 78 & -76.141153 & -0.132021 & 111.8 & 102.9 \\
 & cc-pVDZ & 529 & 276 & -76.241677 & -0.214915 & 116.7 & 107.3 \\

HF & STO-6G & 36 & 21 & -99.526353 & -0.026196 & 100.0 & 99.7 \\
 & DZ & 100 & 55 & -100.146049 & -0.124147 & 107.2 & 100.5 \\
 & cc-pVDZ & 324 & 171 & -100.228652 & -0.209363 & 115.2 & 97.8 \\

NO$^+$ & STO-6G & 64 & 36 & -128.637594 & -0.241971 & 108.7 & 102.4 \\
 & DZ & 256 & 136 & -129.060275 & -0.315068 & 115.3 & 103.6 \\
\end{tabular}
\end{ruledtabular}

\end{table*}

For a variety of molecules and basis sets Table~\ref{t:mol} shows
the percentage of the correlation energy recovered by the
variational 2-RDM method with full-rank and rank-restricted
$N$-representability conditions.  Rank restriction significantly
improves the percentage of correlation energy for all molecules and
basis sets.  For CO the 2-positivity conditions without restriction
yield 108.5\%, 119.3\%, and 121.7\% of the correlation energy in
STO-6G, DZ, and cc-pVDZ basis sets while these condition {\em with}
rank restriction yield 101.2\%, 102.5\%, and 101.1\% of the
correlation energy.  Even though the theoretical rank increases
dramatically with basis-set size, the percentage of the correlation
energy recovered remains nearly constant.  Because the rank
restriction limits the flexibility of the 2-RDM, we observe that the
computed energies with rank restriction are neither consistently
above nor below the FCI energy.

\begin{table*}[htp]

\caption{The percentage of correlation energy along the potential
energy curve of the nitrogen molecule N$_{2}$ in the cc-pVDZ basis
set is reported from the variational 2-RDM method with and without
rank restriction.  At $R=1.485$~\AA\ in a region of the potential
energy curve where the spins are recoupling, sometimes known as the
{\em spin recoupling region}, the error from the rank-restricted
2-RDM method is only +0.002~a.u. relative to FCI.}

\label{t:n2}

\begin{ruledtabular}
\begin{tabular}{ccccc}
\multicolumn{1}{c}{Bond} & \multicolumn{1}{c}{Full CI} &
\multicolumn{1}{c}{Correlation} &
\multicolumn{2}{c}{\% Correlation Energy} \\
\cline{4-5} \multicolumn{1}{c}{Length (\AA)} &
\multicolumn{1}{c}{Energy (a.u.)} & \multicolumn{1}{c}{Energy
(a.u.)} & \multicolumn{1}{c}{Full Rank} &
\multicolumn{1}{c}{Theoretical Rank} \\
\hline
0.80 &  -108.664476 & -0.257118 & 120.7 & 97.5 \\
1.1208 &  -109.282139 & -0.332762 & 120.1 & 98.7 \\
1.175 & -109.275424 & -0.347604 & 120.1 & 98.7 \\
1.475 & -109.141160 & -0.442743 & 119.6 & 99.6 \\
1.85 & -109.008801 & -0.590167 & 116.4 & 97.5 \\
2.225 & -108.970662 & -0.748815 & 111.3 & 96.2 \\
2.6 & -108.963937 & -0.874616 & 108.2 & 95.5 \\
2.975 & -108.962249 & -0.963251 & 106.8 & 95.4 \\

\end{tabular}
\end{ruledtabular}

\end{table*}

Dissociation of the triple-bonded nitrogen molecule N$_{2}$ provides
a classic case of strong electron correlation.  Table~\ref{t:n2} and
Figure~\ref{f:n2} present the potential energy curve of N$_{2}$ in
the cc-pVDZ basis set from the variational 2-RDM method with and
without rank restriction.  At $R=1.485$~\AA, while the 2-RDM method
with the full rank recovers 119.6\% of the correlation energy, the
2-RDM method with the theoretical rank yields 99.6\% of the
correlation energy.  At this distance in a region of the potential
energy curve where the spins are recoupling, sometimes known as the
{\em spin recoupling region}, the error from the rank-restricted
2-RDM method is only +0.002~a.u. relative to FCI.  Figure~\ref{f:n2}
shows that the potential energy curve from the rank-restricted 2-RDM
method closely agrees with the curve from FCI in a large region
surrounding the equilibrium geometry.  The largest errors from rank
restriction occur at significantly stretched geometries where strong
spin entanglement increases the actual rank of the particle-hole
${}^{2} G$ matrix.  In contrast, as observed in previous work, the
2-RDM method without rank restriction has its largest errors in the
spin-recoupling region of the potential energy curve.  One measure
for the potential curve's shape is the {\em non-parallelity error},
the difference between the largest error and the smaller error along
the curve relative to FCI.  While the 2-RDM methods with and without
rank restriction have similar non-parallelity errors over the whole
curve shown in Fig.~\ref{f:n2}, in the region $R \in [0.8,1.85]$ the
rank restriction improves the non-parallelity error from 0.044~a.u.
to 0.013~a.u.

\begin{table*}[htp!]

\caption{The percentage of correlation energy along the potential
energy curve of hydrogen fluoride in the cc-pVDZ basis set is
reported from the variational 2-RDM method with and without rank
restriction. At $R=1.95$~\AA\ in the spin-recoupling region the rank
restriction reduces the error from -44.5~a.u. to -3.1~a.u.}

\label{t:hf}

\begin{ruledtabular}
\begin{tabular}{ccccc}
\multicolumn{1}{c}{Bond} & \multicolumn{1}{c}{Full CI} &
\multicolumn{1}{c}{Correlation} &
\multicolumn{2}{c}{\% Correlation Energy} \\
\cline{4-5} \multicolumn{1}{c}{Length (\AA)} &
\multicolumn{1}{c}{Energy (a.u.)} & \multicolumn{1}{c}{Energy
(a.u.)}
& \multicolumn{1}{c}{Full Rank} & \multicolumn{1}{c}{Theoretical Rank} \\
\hline
0.70 &  -100.129860 & -0.199411 & 115.3 & 96.1 \\
0.9161 &  -100.228633 & -0.209189 & 115.2 & 97.6 \\
1.2 & -100.181953 & -0.222427 & 115.3 & 98.0 \\
1.3 & -100.157836 & -0.227830 & 115.3 & 98.3 \\
1.5 & -100.113798 & -0.240945 & 115.5 & 97.2 \\
1.95 & -100.052917 & -0.283544 & 115.7 & 101.1 \\
2.8 & -100.026420 & -0.369590 & 115.6 & 104.7 \\

\end{tabular}
\end{ruledtabular}

\end{table*}

Due to the high electronegativity of fluorine, the dissociation of
the hydrogen fluoride molecule illustrates the breaking of a polar
covalent single bond.  Table~\ref{t:hf} and Figure~\ref{f:hf}
present the potential energy curve of HF in the cc-pVDZ basis set
from the variational 2-RDM method with and without rank restriction.
At all bond lengths the rank restriction significantly reduces the
error in the percentage of the correlation energy relative to FCI.
At $R=1.95$ in the spin-recoupling region the rank restriction
reduces the error from $-44.5$~a.u. to $-3.1$~a.u.  In the region $R
\in [0.7,2.8]$ shown in Fig.~\ref{f:hf} the rank restriction
decreases the non-parallelity error from 0.027~a.u. to 0.024~a.u.,
and in the region $R \in [0.7,1.95]$ it decreases the
non-parallelity error from 0.014~a.u. to 0.010~a.u.  In contrast to
the triple-bonded N$_{2}$, the singly bonded hydrogen fluoride at
highly stretched geometries has energies from the rank restriction
below those from FCI.

\section{Discussion and Conclusions}

\label{sec:con}

Variational minimization of the ground-state energy as a function of
the 2-RDM~\cite{ME01,NNE01,M02,ZBF04,M04c,M04b,CSL06,M06b,M06c,M07b,
NBF08,BPZ07,GM08a,VVV09,GM10b,SI10}, constrained by
$N$-representability conditions, provides a polynomial-scaling
approach to studying strongly correlated molecules without computing
the many-electron wavefunction.  Here we have introduced a new
approach to enhancing necessary conditions for $N$-representability
through rank restriction of the 2-RDM. Applications were made to
molecules at both equilibrium and non-equilibrium geometries.

An important set of $N$-representability conditions on the 2-RDM is
the 2-positivity conditions, which restrict the probability
distributions of two particles ($^{2} D$), two holes ($^{2} Q$), and
a particle-hole pair ($^{2} G$) to be nonnegative.  In
section~\ref{sec:wf} we derived the maximum rank of the
particle-hole ${}^{2} G$ metric matrix for two model wavefunctions,
the Hartree-Fock and the AGP wavefunctions.  The Hartree-Fock wave
functions are a small subset of the AGP wavefunctions, and hence,
their particle-hole matrices have a maximum rank $(r-N)N+1$ which is
strictly less than the maximum rank $r(r-1)/2$ of the AGP
particle-hole matrices.  Because the 2-positivity conditions
constrain AGP Hamiltonians---that is, Hamiltonians with AGP
ground-state wavefunctions---to yield the exact ground-state
energies, the rank of the AGP $^{2} G$ matrix provides a minimum
rank for the molecular particle-hole $^{2} G$ matrix within
variational 2-RDM calculations of general systems.  Selecting a
smaller rank for $^{2} G$ would render the variational 2-RDM method
inexact for AGP Hamiltonians. Unlike the case in Hartree-Fock
theory, the maximum rank of $^{2} G$ within the AGP ansatz is
independent of the number $N$ of particles, which reflects its
independence from a reference determinant wavefunction and hence,
its ability to treat strong electron correlation.

The variational 2-RDM method with rank-restricted 2-positivity
conditions was applied to computing the energies and 2-RDMs for a
variety of molecules at equilibrium geometries as well the potential
energy curves of the nitrogen and hydrogen fluoride molecules.
Specifically, the rank of the singlet spin block of the
particle-hole matrix was restricted to its maximum value from an AGP
wavefunction $r_{s}(r_{s}+1)/2$ with $r_{s}=r/2$. The
rank-restricted conditions were implemented through an extension of
the first-order matrix-factorization algorithm for large-scale SDP.
The results demonstrate that rank restriction significantly improves
the accuracy of computed energies. For example, the percentages of
correlation energies recovered for HF, CO, and N$_{2}$ improve from
115.2\%, 121.7\%, and 121.5\% without rank restriction to 97.8\%,
101.1\%, and 100.0\% with rank restriction, respectively. The
improvement occurs at equilibrium and non-equilibrium geometries and
across basis sets.  Computationally, the rank-restricted conditions
are slightly less expensive than the full 2-positivity conditions.
Rank restriction removes degrees of freedom that are not
sufficiently constrained by the 2-positivity conditions without
sacrificing the method's ability to treat strong electron
correlation, as seen in the bond dissociation of N$_{2}$. Although
further research is needed to study the method in larger molecules
such as polyaromatic hydrocarbons~\cite{GM08a,PGG11} and firefly
luciferin~\cite{GM10b}, the present results indicate that rank
restriction is a promising approach to improving the 2-positivity
conditions within the variational 2-RDM method without increasing
computational cost.

\begin{acknowledgments}

DAM gratefully acknowledges the NSF, ARO, Microsoft Corporation,
Dreyfus Foundation, and David-Lucile Packard Foundation for their
support.

\end{acknowledgments}

\bibliography{RDM_Rank_BIB_V1}

\end{document}